\documentclass[aps,twocolumn,prb,showpacs,superscriptaddress,10pt,floatfix]{revtex4-1}
\usepackage{amssymb}
\usepackage{amsmath}
\usepackage{graphicx}
\usepackage{dcolumn}
\usepackage{bm}
\usepackage[usenames]{color}
\usepackage{textcomp} % me: for \textmu

\usepackage[implicit=true,colorlinks=true,linkcolor=blue,citecolor=blue,urlcolor=blue]{hyperref} % implicit: no eq,sec... as hyperref

\DeclareMathOperator{\Real}{\mathrm{Re}}

\newcommand{\units}[1]{\ensuremath{\,\mathrm{#1}}}

\newif\ifcom
\newif\ifdel

\newcommand{\PartDeriv}[2]{\frac{\partial #1}{\partial #2}}

\newcommand{\ii}{\mathrm{i}}
\newcommand{\ee}{\mathrm{e}}

\newcommand{\Omegat}{\widetilde{\Omega}}

\usepackage[normalem]{ulem}
\definecolor{uniorange}{RGB}{250, 100, 55}
\definecolor{darkgreen}{rgb}{0.2,.7,0.2}
\usepackage{xcolor,cancel} % to strike out equations

\begin{document}

\title{The effect of normal metal layers in ferromagnetic Josephson junctions}

\author{D.\,M.~Heim}
\email{dennis.heim@uni-ulm.de}
\affiliation{Institut f{\"u}r Quantenphysik and Center for Integrated Quantum Science and Technology (IQ$^{ST}$), Universit{\"a}t Ulm, D-89069 Ulm, Germany}

\author{N.\,G.~Pugach}
\affiliation{Skobeltsyn Institute of Nuclear Physics, M.~V. Lomonosov Moscow State University, 119991 Leninskie Gory, Moscow, Russia}
\affiliation{Royal Holloway University of London, Egham, Surrey, TW20 0EX, United Kingdom}

\author{M.\,Yu.~Kupriyanov}
\affiliation{Skobeltsyn Institute of Nuclear Physics, M.~V. Lomonosov Moscow State University, 119991 Leninskie Gory, Moscow, Russia}
\affiliation{Moscow Institute of Physics and Technology, Dolgoprudny, Moscow Region, Russia}

\author{E.~Goldobin}
\author{D.~Koelle}
\author{R.~Kleiner}
\affiliation{Physikalisches Institut and Center for Collective Quantum Phenomena in LISA$^+$, Universit\"{a}t T\"{u}bingen, D-72076 T\"{u}bingen, Germany}

\author{N.~Ruppelt}
\affiliation{Nanoelektronik, Technical Faculty, University of Kiel, D-24143 Kiel, Germany}

\author{M.~Weides}
\affiliation{Physikalisches Institut, Karlsruhe Institute of Technology, D-76131 Karlsruhe, Germany}

\author{H.~Kohlstedt}
\affiliation{Nanoelektronik, Technical Faculty, University of Kiel, D-24143 Kiel, Germany}

\begin{abstract}
Using the Usadel equation approach, we provide a compact formalism to calculate the critical current density of 21 different types of ferromagnetic (F) Josephson junctions containing insulating (I) and normal metal (N) layers in the weak link regions. In particular, we obtain that even a thin additional N layer may shift the $0$-$\pi$ transitions to larger or smaller values of the thickness $d_F$ of the ferromagnet{, depending on its conducting properties}. For certain values of $d_F$, a $0$-$\pi$ transition can even be achieved by changing only the N layer thickness. We use our model to fit experimental data of SIFS and SINFS tunnel junctions, where S is a superconducting electrode.
\end{abstract}


\maketitle

\section{Introduction}\label{sec:introduction}

The coexistence and competition of ferromagnetic and superconducting ordering leads to a rich spectrum of unusual physical phenomena, intensively studied during the recent years.\cite{golubov:2004,Buzdin:2005:Review,VolkovRevModPhys} One of the consequences is the so-called $\pi$ Josephson junction with phase shift $\pi$ in the ground state. This development makes the ferromagnetic Josephson junctions (FJJs) a subject of intensive theoretical and experimental studies.

\begin{figure}[ht!]
% \begin{center}
\includegraphics{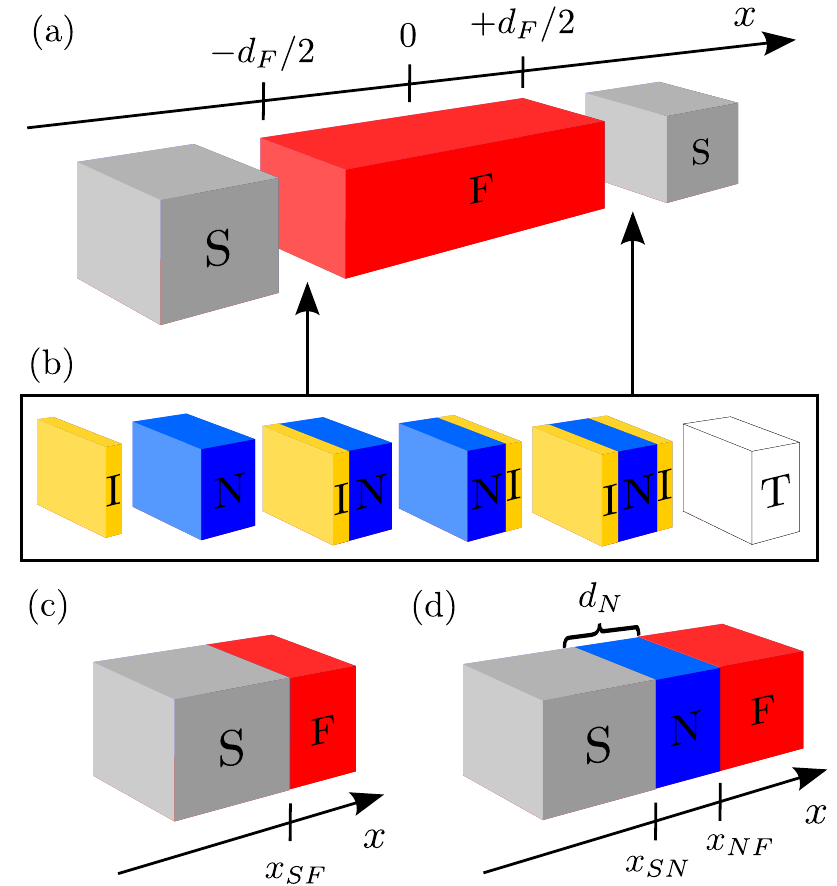}
  \caption{{The FJJ configurations we consider. In (a) we show the basic geometry consisting of two thick superconducting (S) electrodes separated by  a ferromagnetic (F) weak link of thickness $d_F$. Our formalism covers all 21 FJJs resulting from the insertion of each of the layers shown in (b) at the SF interfaces in (a). These layers are composed of insulating (I) or normal metal (N) films. The case of no additional layer is denoted by T (transparent interface). In (c) and (d) we define parameters for the derivation of our formalism.}}
  \label{fig:junction_02}
% \end{center}
\end{figure}

An FJJ usually contains two thick superconducting (S) electrodes with a ferromagnetic (F) film between them{, see Fig.~\ref{fig:junction_02}(a)}. In the present article we derive a formalism to calculate the critical current densities of FJJs with additional insulating (I) and normal metal (N) layers at the SF interfaces.
{A sign reversal of the critical current indicates a transition from the 0 to the $\pi$ state of the junction. This transition is usually realized by changing the thickness $d_F$ of the ferromagnet. However, we show that it can also be achieved by only changing the thickness $d_N$ of an N layer when inserted at the SF interface in SINF configuration.}

{The geometry of all} FJJs {we consider} can be constructed by selecting one of the items of Fig.~\ref{fig:junction_02}(b) and inserting it by following one of the arrows into Fig.~\ref{fig:junction_02}(a). At the other arrow position we insert either the same or another item from Fig.~\ref{fig:junction_02}(b). In this way we obtain 21 possible FJJ configurations.

{The purpose of the additional I layer(s) is to} enlarge the product $J_c R_N$ in the $\pi$-state. Here $J_c$ is the critical current density of the junction and  $R_N$ is its normal resistance, which is mainly determined by the insulating barrier(s).

One reason why we consider additional N layers is the existence of a so-called ``dead'' layer which is assumed in order to fit many experiments.\cite{RyazanovBuzdinJc(d)PRL2006,Blum:2002:IcOscillations,Sellier:2003:SFS,Weides:2006:SIFS-HiJcPiJJ,Weides:0-piLJJ,Pfeiffer:2008:SFS:semifluxon,Weides:2009} This dead layer is a part of the ferromagnet which behaves as a nonmagnetic metal. It usually appears due to the surface roughness or the mutual dissolution of N and F layers. It is inherent for example for Cu, which is very popular as a spacer, and its alloys with 3d metals. Usually it is assumed that the dead layer makes the effective F layer thinner.
However, does the dead layer cause additional effects, and at what conditions? Our calculations aim to answer this question. 

The answer to this question is also important for the development of magnetic memory cells for rapid single flux quantum (RSFQ) logics which becomes more and more actual.\cite{goldobin:2013:memory,qader:2014,robinson:2014,niedzielski:2014,alidoust:2014,baek:2014,iovan:2014} Only recently, a new type of magnetic memory element based on FJJs with a complex insulator-superconductor-ferromagnet weak link (SIsFS) was proposed.\cite{Ryazanov:2013:SIsFS-theor,Ryazanov:2012:SIsFS-exper} These FJJs have a large $J_c R_N$ product in the $\pi$-phase. The middle superconducting ``s'' layer is inserted in the weak link to recover the superconducting pairing and increase $J_c$. The thickness of this layer is of the order of the coherence length so that it may make a transition to the normal state at different conditions than the thick outer S electrodes. One of the aims of our calculation is to study the behavior of such SIsFS FJJs when their middle superconducting layer is in the normal state.

Ferromagnetic Josephson junctions can also be used as (non-dischargeable) on-chip $\pi$-phase batteries for self-biasing various electronic circuits in classical and quantum domains, e.g.~self-biased RSFQ logic\cite{Ortlepp:2006:RSFQ-0-pi} or flux qubits.\cite{Ioffe:1999:sds-waveQubit,OurQubit,Yamashita:2006:pi-qubit:3JJ} In classical circuits, a phase battery may also substitute the conventional inductance and substantially reduce the size of an elementary cell.\cite{Ustinov:2003:RSFQ+pi-shifters} Some of these proposals were already realized practically.\cite{Ortlepp:2006:RSFQ-0-pi,Ryazanov:2010} The key question for their realization is the range of parameters, e.g. the ferromagnetic layer thickness $d_F$, at which the $\pi$ ground state is established, that is, the $0$-$\pi$ transition occurs. In recent works \cite{Vasenko,Buzdin:PiJJ:JETPL2003} it was shown that the presence of extra insulating layers shifts the first $0$-$\pi$ transition to smaller values of $d_F$. The explanation of this effect is that the order parameter decreases step-wise at the I barrier(s) so that one requires a thinner F layer to reach the $0$-$\pi$ transition.

Introducing an N layer between the ferromagnet and the S electrode {was} technologically necessary {in many FJJ experiments}.\cite{Ryazanov:2001:SFS-PiJJ,RyazanovBuzdinJc(d)PRL2006,Blum:2002:IcOscillations,Sellier:2003:SFS,Born:2006:SIFS-Ni3Al,Weides:2006:SIFS-HiJcPiJJ,Weides:0-piLJJ,Pfeiffer:2008:SFS:semifluxon,robinson:2006,RobinsonPRB76,Gross2010} However, such a situation was not taken into account by any theoretical explanation of these experiments, {or considered in previous theoretical works on tunnel FJJs both in the clean and dirty limits,\cite{golubov:2002,golubov:2005,OurPRB2009,liu:2010,Vasenko:2010,vasenko:2011} see Refs.~\onlinecite{Buzdin:2005:Review,golubov:2004} for review.} As we show in the current paper, this is only reasonable if the F and N metals behave fully identically, except for their magnetic properties. Otherwise, the presence of a thin N layer changes the boundary conditions which influences, in particular, the dependence of the Josephson current density $J_c$ on the F layer thickness $d_F$. Recent experiments,\cite{Ryazanov2harmNovgorod} which use a new continuous in-situ technology allowing the deletion of this layer, exhibit actually a change of the $0$-$\pi$ transitions in the $J_c (d_F)$
dependence.

The article is organized as follows. In Sec.~\ref{sec:model} we describe our model based on the Usadel equations supplemented with Kupriyanov-Lukichev boundary conditions. Different types of interlayer boundaries are analyzed. Section \ref{sec:discussion} presents the obtained dependencies of the critical current density on the F layer thickness as well as the analysis of the $0$-$\pi$ transitions in the framework of a linear approximation. {We use our formalism in Sec.~\ref{sec:comparison} to fit experimental data of SINFS and SIFS junctions.} Section \ref{sec:conclusion} concludes this work. Details of the calculation can be found in the appendix.

\section{Model}\label{sec:model}

\subsection{The boundary value problem}\label{sec:BVP}

{The basic Josephson junction configuration} we consider is sketched in Fig.~\ref{fig:junction_02}{(a)}. It consists of two thick S electrodes enclosing an F layer {of the thickness $d_F$} along the $x$ axis. Our model {allows to consider} an additional {I or} N layer at the SF interfaces as well as I layers at the SN or NF interfaces{, as illustrated by Fig.~\ref{fig:junction_02}(b).}

We calculate the critical current density $J_c$ of these configurations by determining their Green's functions in the ``dirty'' limit. In this limit, the elastic electron scattering length is much smaller than the characteristic decay length of the superconducting wave function. We determine the Green's functions with the help of the Usadel equations,~\cite{usadel:1970} which we use similar to Ref.~\onlinecite{Buzdin:2005:Review} in the form
\begin{align}
  &{\xi_j^2} \left( G_j  \frac{\partial^2}{\partial x^2} F_j
  - F_j  \frac{\partial^2}{\partial x^2} G_j\right)
  - \left( \widetilde\Omega {+\eta\, G_j} \right) F_j = 0, \nonumber \\
  &G_j^2+F_j\widetilde{F}_j=1, \quad j \in \{{N,F}\},
  \label{eq:usadel:general}
\end{align}
in the N and F layer, where $F_j$ and $G_j$ are the Usadel Green's functions, while $\widetilde{F}_j(\omega) \equiv F_j^* (-\omega)$. The frequencies $\Omegat \equiv \Omega+\ii h$ contain the scaled Matsubara frequencies $\Omega \equiv {\omega}/{(\pi T_c)}$, where $\omega \equiv \pi T (2n+1)$ at the temperature $T$, and $T_c$ is the critical temperature of the superconductor. {By using the definition $\eta \equiv 1/(\tau_m\pi T_c)$ we take, similar to} {Ref.} \onlinecite{Vasenko}, {the spin-flip scattering time $\tau_m$ into account. This approach requires a ferromagnet with strong uniaxial anisotropy} {like for example Cu alloys with transition metals, which are used in many experiments.} Equation (\ref{eq:usadel:general}) should be satisfied for any integer number $n$. The scaled exchange energy $h \equiv {H}/{(\pi T_c)}$ of the ferromagnetic material, where the energy $H$ describes the {exchange integral of the conducting electrons}, is assumed to be zero in the N layer.

{In our model we use} the coherence lengths
\begin{equation}
  \xi_N \equiv \sqrt{\frac{D_{N}}{2\pi T_c}}, \quad \xi_F \equiv \sqrt{\frac{D_{F}}{2\pi T_c}} {, \quad \xi_H \equiv \sqrt{\frac{D_F }{ H}}}
  \label{eq:def:xi}
\end{equation}
of the superconducting correlations{, which} are defined with the help of the diffusion coefficients $D_{N}$ and $D_{F}$ in the normal and ferromagnetic metal, respectively. We use the scaling defined by $\hbar \equiv k_{B} \equiv 1$.

{The decay length $\xi_H$ of superconducting correlations in the ferromagnet is usually in the order of nm. Therefore this is sufficiently small ($\xi_H \lesssim d_F$) to consider} the supercurrent as a result of interference of anomalous Green's functions induced from the superconducting banks. It is convenient to consider this problem in theta parametrization\cite{zaikin:1981}
\begin{equation}
 F_j=\ee^{\ii \varphi_j} \sin \theta_j,\quad G_j=\cos \theta_j
 , \label{eq:theta:param}
\end{equation}
where $\varphi_j$ is independent of the coordinate $x$. It corresponds to the phase $\varphi_j \equiv \pm \phi/2$ of the order parameter of the S banks for the right and left superconducting electrode respectively, while  $\theta_j$ satisfies the sine-Gordon type differential equation
\begin{equation}
  \xi_j^2\ \frac{\partial^2}{\partial x^2} \theta_j - \left( \widetilde{\Omega} + { \eta \cos \theta_j} \right)\sin \theta_j= 0
  . \label{eq:usadel:theta}
\end{equation}

Since we assume that the superconductivity in the S electrodes is not suppressed by the neighboring N and F layers, we obtain
\begin{equation}
 \theta_{S}=\arctan \frac{\Delta}{\omega}
    \label{eq:theta_S}
\end{equation}
{analogous to Vasenko \emph{et al.}\cite{Vasenko}} at the interfaces of the superconductor, where $\Delta$ is the absolute value of the order parameter in the superconductor. The validity of this assumption depends on the values of the suppression parameters
\begin{align}
 \gamma_{BSF}& \equiv \frac{R_{BSF}A_{BSF}}{\rho_{F}\xi_{F}},\quad \gamma_{SF} \equiv \frac{\rho_{S}\xi_{S}}{\rho_{F}\xi_{F}}, \nonumber \\
 \gamma_{BSN}& \equiv \frac{R_{BSN}A_{BSN}}{\rho_{N}\xi_{N}}, \quad \gamma_{SN} \equiv \frac{\rho_{S}\xi_{S}}{\rho_{N}\xi_{N}}
  \label{eq:gamma:bsn}
\end{align}
at the S boundaries, which we discuss in more detail in Subsec. \ref{sec:Interfaces}. Here we use the resistances $R_{BSF}$, $R_{BSN}$ and the areas $A_{BSF}$, $A_{BSN}$ of the SN and SF interfaces. The values $\rho_{N}$, $\rho_{F}$ and $\rho_{S}$ describe the resistivity of the N, F, and S metals, respectively.

The Kupriyanov-Lukichev boundary condition~\cite{kupriyanov:1988,koshina:2000} at the superconducting interface, shown in {Fig.~\ref{fig:junction_02}(c), is}
  \begin{align}
    {
    \sin(\theta_{{F,S}}-\theta_S) = \gamma_{BSF}\ \xi_F \left[ \PartDeriv{}{x} \theta_F \right]_{x_{SF}}
    }
    , \label{eq:boundary:fs}
  \end{align}
{where $\theta_{{F,S}} \equiv \theta_F(x_{SF})$, while in Fig.~\ref{fig:junction_02}(d) it is}
  \begin{align}
   \sin(\theta_{{N,S}}-\theta_S) = \gamma_{BSN}\ \xi_N \left[ \PartDeriv{}{x} \theta_N \right]_{x_{SN}}
   ,  \label{eq:boundary:sn}
  \end{align}
{where $\theta_{{N,S}} \equiv \theta_N(x_{SN})$, at the SN boundary and}
\begin{align}
    {
    \sin(\theta_{F,N}-\theta_{N,F}) = \gamma_{BNF}\ \xi_F \left[ \PartDeriv{}{x} \theta_F \right]_{x_{NF}}
    }
    \label{eq:boundary:nf:disc}
\end{align}
{at the NF boundary. Here we defined $\theta_{{F,N}} \equiv \theta_F(x_{NF})$ and $\theta_{{N,F}} \equiv \theta_N(x_{NF})$.  Additionally we use } the differentiability condition
  \begin{eqnarray}
    \gamma_{NF}\ \xi_F \left[ \PartDeriv{}{x} \theta_F \right]_{x_{NF}} = \xi_N \left[ \PartDeriv{}{x} \theta_N \right]_{x_{NF}}
    . \label{eq:boundary:nf:diff}
  \end{eqnarray}
The suppression parameters
\begin{align}
 {\gamma_{BNF} \equiv \frac{R_{BNF}A_{BNF}}{\rho_{F}\xi_{F}}}
 ,\quad \gamma_{NF} \equiv \frac{\rho_{N}\xi_{N}}{\rho_{F}\xi_{F}}
 \label{eq:gamma:bnf}
\end{align}
are defined analogous to Eq.~\eqref{eq:gamma:bsn}, but not restricted to only small or large values.

In order to finally extract the critical current density
$J_c$ from the current phase relation $J(\phi)=J_c \sin \phi$ we will calculate the total current density~\cite{golubov:2004}
\begin{align}
 J(\phi)=\ii \frac{\pi T}{2 e \rho_F} \sum_{\omega=-\infty}^{\infty}&
   \left[ F_F (\omega)\ \PartDeriv{}{x} F_F^*(-\omega) \right. \nonumber \\
   &\left. - F_F^*(-\omega)\ \PartDeriv{}{x} F_F(\omega) \right]_{x=0}
  \label{eq:current:f:tot:def}
\end{align}
flowing through our device, with the help of the Green's function $F_F$ in the F layer. Here we chose the position $x=0${, see Fig.~\ref{fig:junction_02}(a),} in order to simplify the calculation.

\subsection{Critical current density}\label{sec:calculation}

In this section we rewrite expression \eqref{eq:current:f:tot:def} to be able to directly calculate the critical current densities of all SFS Josephson junctions {of the type sketched in Fig.~\ref{fig:junction_02}(a), which may include each one of the} {layers shown in Fig.~\ref{fig:junction_02}(b) at the} SF interfaces. 

In order to solve the Usadel equations \eqref{eq:usadel:general} in the F layer we use the ansatz \cite{buzdin:1991,Vasenko}
\begin{equation}
 F_F(x)=\ee^{-\ii \phi/2} \sin[\theta_F^-(x)] + \ee^{+\ii \phi/2} \sin[\theta_F^+(x)]
 , \label{eq:ansatz:f}
\end{equation}
where each function $\theta_F^-(x)$ and $\theta_F^+(x)$ solves the non-linear differential equation \eqref{eq:usadel:theta} for $j={F}$. Additionally we use the conditions $\theta_F^\pm=0$ and $\partial {\theta_F^\pm}/\partial x =0$ at $x=\mp\infty$. Then the solution $\theta_F^-(x)$ will turn out to be most dominant in the left side of the F part and to decay exponentially in the right side of the junction. Therefore, it has practically no overlap with the solution $\theta_F^+(x)$ which is dominant in the right side of the F layer. It was shown \cite{Vasenko} that this ansatz is valid even for small distances $d_F \sim \xi_H$, that is, in the region of the first $0$-$\pi$ transition, where $\xi_H$ {is defined by Eq.~\eqref{eq:def:xi}}.

We obtain both solutions $\theta_F^-(x)$ and $\theta_F^+(x)$ by integrating the differential equation \eqref{eq:usadel:theta} for $j={F}$ twice. The first integration results in
  \begin{equation}
   \PartDeriv{}{x} \theta_F^\pm= \pm \frac{2}{\xi_F} \sqrt{\widetilde\Omega {+\eta \cos^2\frac{\theta_F^\pm}{2}}} \sin \frac{\theta_F^\pm}{2}
   , \label{eq:usadel:theta:integrated}
  \end{equation}
 {where $\theta_F^\pm \equiv \theta_F^\pm(x)$.}
A second integration leads us {by using the definition $q \equiv \sqrt{\widetilde\Omega + \eta}$} to the equation\cite{Vasenko,faure:2006}
  \begin{align}
   {\frac{\sqrt{\widetilde\Omega+\eta \cos^2 \frac{\theta_F^\pm}{2}}-q \cos \frac{\theta_F^\pm}{2}}
   {\sqrt{\widetilde\Omega+\eta \cos^2\frac{\theta_F^\pm}{2}}+q \cos \frac{\theta_F^\pm}{2}}
   = g^\pm \exp \left[ \pm \frac{2q}{\xi_F} \left(x \mp \frac{d_F}{2}\right) \right] }
   . \label{eq:sol:f:r:pre}
  \end{align}
  {Here $g^\pm$ are the integration constants. In the F layer we can assume small superconducting correlations $\theta_F \ll 1$ to linearize the denominator of the left-hand side of Eq.~\eqref{eq:sol:f:r:pre} which leads us to the equation}
  \begin{align}
  {
    \sin \frac{\theta_F^\pm(x)}{2}= \chi^\pm \exp \left[ \pm\frac{{q}}{\xi_F}
   \left( x \mp \frac{d_F}{2} \right) \right]
   }
   . \label{eq:sol:f:r}
  \end{align}
  The {rewritten integration} constants {$\chi^\pm$} are given by the {boundary conditions at the right and left ferromagnetic interfaces} as
  \begin{align}
   {
   \chi^+ \equiv \sin \frac{\theta_{F}(+d_F/2)}{2}, \quad \chi^- \equiv \sin \frac{\theta_{F}(-d_F/2)}{2}
   }
   . \label{eq:g:def:r}
  \end{align}

By inserting our ansatz \eqref{eq:ansatz:f} with the solutions \eqref{eq:sol:f:r} into the current density \eqref{eq:current:f:tot:def} and by using the approximation $\Omegat \approx \ii h$, which holds for the condition $\pi T_c \ll H$, and the assumption {$\xi_H \lesssim d_F$}, we obtain the critical current density
   \begin{align}
  {
  J_c = 16 \frac{\pi T}{e \rho_{F}} \sum_{\omega>0}\ \mathrm{Re}
  \left( \gamma\ \ee^{-\gamma d_F} \chi^+\, \chi^- \right),\quad \gamma = \frac{q}{\xi_F}
  }
  . \label{eq:current:f:tot:res}
 \end{align}
{The constants $\chi^\pm$} will be determined in the next section.

\subsection{{SF interface without or including an N layer}} \label{sec:Interfaces}

{In the following we determine} a constant {$\chi_{TI}$ to replace $\chi^+$ or $\chi^-$ in Eq.~\eqref{eq:current:f:tot:res} in the case of no N layer at an SF interface, as shown for example in} Fig.~\ref{fig:junction_02}{(c). The index TI stands for transparent or insulating.} 

We insert the integrated sine-Gordon equation \eqref{eq:usadel:theta:integrated} at the position $x_{SF}$ into the boundary condition \eqref{eq:boundary:fs} and obtain the relation
  \begin{align}
    2 \gamma_{BSF} {\sqrt{\widetilde\Omega {+\eta \cos^2\frac{\theta_{{F,S}}}{2}}}}\ \sin \frac{\theta_{{F,S}}}{2}
      = \sin(\theta_S-\theta_{{F,S}})
   . \label{eq:boundary:fs:modified}
  \end{align}
{By defining $\chi_{TI} \equiv \sin({\theta_{F,S}}/{2})$ analogous to Eq.~\eqref{eq:g:def:r}} we rewrite Eq.~\eqref{eq:boundary:fs:modified} {in the form}
  \begin{align}
   \chi_{TI}^4 & + 2 \gamma_{BSF} \sqrt{\widetilde\Omega {+ \eta(1-\chi_{TI}^2)}} \sin\theta_S\ \chi_{TI}^3  \nonumber \\
    &+ \{\gamma_{BSF}^2 [ \widetilde\Omega {+ \eta(1-\chi_{TI}^2)} ] - 1\}\ \chi_{TI}^2 \nonumber \\
   &-\gamma_{BSF} \sqrt{\widetilde\Omega {+ \eta(1-\chi_{TI}^2)}} \sin\theta_S\ \chi_{TI} \nonumber \\
   &+ \frac{1}{4}{\sin^2\theta_S} = 0
   . \label{eq:chi_r}
  \end{align}
{In the case $\eta \rightarrow 0$,} this equation is {a quartic equation in $\chi_{TI}$ and therefore} exactly solvable. 
{To find the solutions in this case we use the function \emph{solve} of the software MATLAB.
Afterwards we make use of Eq.~\eqref{eq:boundary:fs:modified} to select one of the four solutions.} {In the case $\eta \neq 0$ we solve Eq.~\eqref{eq:chi_r} numerically} {by using the function \emph{fsolve} of the software MATLAB together} {with the solution of the limit $\eta \rightarrow 0$ as starting value.}

In this way we find {$\chi_{TI}$} for the determination of the critical current density \eqref{eq:current:f:tot:res} {in the case of no N layer at the SF boundary}. The case of a small parameter $\gamma_{BSF}$ corresponds to a transparent SF interface{, while a large one corresponds to an insulating interface.\cite{Vasenko,faure:2006}} 

Next we determine a constant {$\chi_N$ for} the case of a thin N layer $d_N\ll\xi_N$ between the superconductor and ferromagnet as shown in Fig.~\ref{fig:junction_02}{(d)}.

{By inserting the integrated sine-Gordon equation~\eqref{eq:usadel:theta:integrated} for $x=x_{NF}$ into the boundary condition \eqref{eq:boundary:nf:disc} we obtain the equation}
  \begin{align}
   {    2 \gamma_{BNF}  {\sqrt{\widetilde\Omega +\eta \cos^2\frac{\theta_{F,N}}{2}}}\ \sin \frac{\theta_{F,N}}{2}
   = \sin(\theta_{N,F}-\theta_{F,N})
    \label{eq:boundary:nf:modified}
   .}
  \end{align}
{When we rewrite this equation using} the definition {${\chi_N} \equiv \sin({\theta_{F,N}}/2)$}{, the result}
  \begin{align}
     { {\chi_N}^4 } & {+ 2 \gamma_{BNF} \sqrt{\widetilde\Omega + \eta(1-\chi_N^2)} \sin{\theta_{N,F}}\ {\chi_N}^3 } \nonumber \\
     & { + \{\gamma_{BNF}^2[\widetilde\Omega + \eta(1-\chi_N^2)] - 1\}\ {\chi_N}^2} \nonumber \\
   & { -\gamma_{BNF} \sqrt{\widetilde\Omega + \eta(1-\chi_N^2)} \sin{\theta_{N,F}}\ {\chi_N} } \nonumber \\
   & { + \frac{1}{4} {\sin^2{\theta_{N,F}}}  = 0 }
   \label{eq:chi_l}
  \end{align}
looks similar to Eq.~\eqref{eq:chi_r}. The main difference is that it {reduces in the case $\eta \rightarrow 0$} not to an equation of fourth order in {$\chi_N$.} This is because we take the inverse proximity effect at the NF boundary into account. Therefore, the value $\theta_{{{N,F}}}${$\equiv \theta_N(x_{NF})$} depends also on $\chi_N$, which itself is related to {$\theta_{F,N}\equiv \theta_F(x_{NF})$}, even in the case $d_N\ll \xi_N$, as we show in the appendix.

{
However, we also show in the appendix that Eq.~\eqref{eq:chi_l} reduces in the limit $\eta \rightarrow 0$ together with $\gamma_{NF} \rightarrow 0$ to an equation of fourth order in $\chi_N$. Therefore, we make three steps in order to solve Eq.~\eqref{eq:chi_l}. First we determine its solution in the case $\eta,\gamma_{NF} \rightarrow 0$ similar to the forth-order case of Eq.~\eqref{eq:chi_r}. We then use this result as a starting value to solve Eq.~\eqref{eq:chi_l} for only the limit $\eta \rightarrow 0$ with the help of the function \emph{fsolve} of the software MATLAB. This in turn leads to another starting value which we use to solve Eq.~\eqref{eq:chi_l} with \emph{fsolve}, but without any limiting case.
}

{The solution $\chi_N$ of Eq.~\eqref{eq:chi_l} finally can be used as $\chi^+$ or $\chi^-$ for the determination of the critical current density \eqref{eq:current:f:tot:res} in the case of an N layer at the SF interfaces. Small parameters $\gamma_{BSN}$ and $\gamma_{BNF}$ correspond to transparent SN and NF interfaces{, while large ones correspond to insulating interfaces.\cite{Vasenko,faure:2006}}}

\section{Discussion}\label{sec:discussion}

{In this section} we first select {FJJ} configurations, where the N layer has the largest influence. {We then analyze their critical current densities with the help of the formalism we derived in the previous section. Finally we} discuss the results with the help of solutions of the linearized differential equation~\eqref{eq:usadel:theta}.

We do not analyze configurations, where a thin N layer ($d_N\ll\xi_N$) is located between S and I layers, which gives only a negligible reduction of $J_c$ compared to the case without an N layer. This is because the superconducting condensate just penetrates into the whole N layer. The same effect occurs when the thin N spacer separates the S and F layers and both (SN and NF) interfaces are transparent.

\begin{figure*}[!t]
\begin{center}
\includegraphics{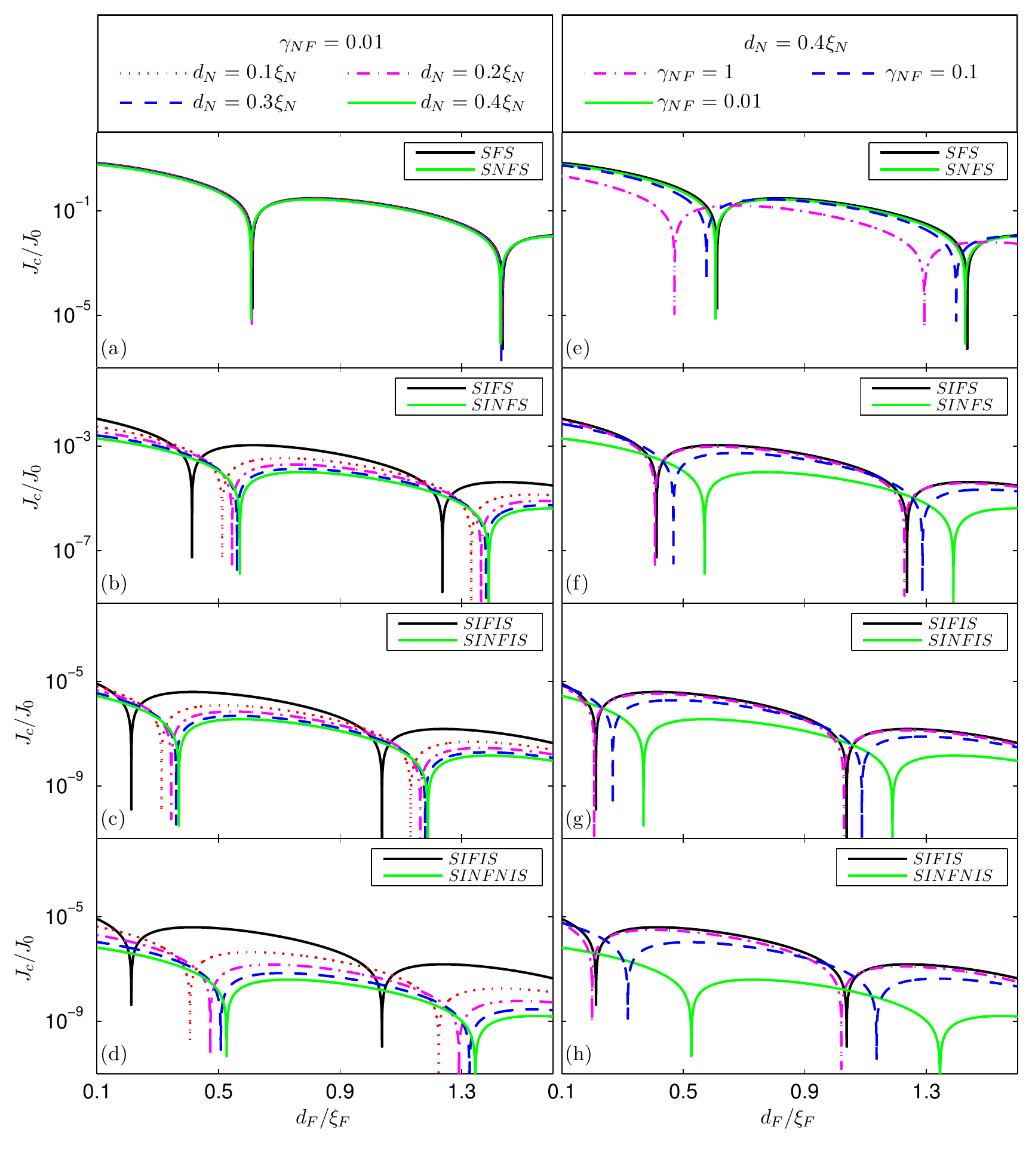}
  \caption{The critical current density $J_c(d_F)$ calculated using Eq.~\eqref{eq:current:f:tot:res} for different FJJs in units of $J_0={\pi T_c}/{(\rho_F \xi_F e)}$. {Colored} lines {correspond to SFS junctions including N layers.} The {solid black} lines are solutions without N layer {and} {in agreement with Refs.~}\onlinecite{buzdin:1991,Vasenko}. {The} {colored} {lines} in Figs. (a--d) {are dotted for $d_N=0.1\xi_N$, dashed-dotted for $d_N=0.2\xi_N$, dashed for $d_N=0.3\xi_N$, and solid for $d_N=0.4\xi_N$.} Here we used the suppression parameter $\gamma_{NF}=0.01$. In Figs. (e--h) the {dashed-dotted} {lines correspond to $\gamma_{NF}=1$, the} {solid lines to $\gamma_{NF}=0.1$, and the } {dashed lines to $\gamma_{NF}=0.01$} {at} the
fixed thickness $d_N=0.4 \xi_N$. {We used the suppression parameters given by Tab.~\ref{tab:gamma_B}.}
Additionally we chose $h=30$, $T_c=9.2\units{K}$, $T=0. 5T_c$ and $\eta=0$.
From Figs.~(b-d) we conclude  that inserting an N layer can mitigate the effect of the insertion of an I layer, and Figs. (f-h) show that this behavior depends strongly on $\gamma_{NF}$.}
  \label{fig:currents_plot_gNFchg}
\end{center}
\end{figure*}

However, when the SN boundary has a very weak transparency or gets even insulating, that is, the N layer is located between an I and F layer, then the N layer(s) play(s) a more notable role depending on the relation of resistances $\gamma_{NF}$ (\ref{eq:gamma:bnf}), as we will see in the following.

Examples for the critical current density $J_c (d_F)$ in these situations are presented in Fig.~\ref{fig:currents_plot_gNFchg} with different numbers of insulating barriers. To in- and exclude these barriers we use the boundary parameters shown in Tab.~\ref{tab:gamma_B}. Since we only want to change N-layer properties, like $d_N$, $\rho_N$ or $\xi_N$ of the same junction, we keep the product
\begin{align}
{
 \gamma_{BSN}\gamma_{NF} = \frac{R_{BSN}A_{BSN}}{\rho_F \xi_F}
 \label{eq:product:gamma}
 }
\end{align}
constant.

Each figure~\ref{fig:currents_plot_gNFchg}(a-d) shows several dependences $J_c(d_F)$ for FJJs containing N layers of different thicknesses and the corresponding reference FJJ without any N layer ({solid black} lines\cite{Vasenko,BuzdinBaladie2003}). {The I layers in all panels of Fig.~\ref{fig:currents_plot_gNFchg}(b-d,f-h) are chosen to be exactly identical.} Here we observe that the additional N layer at the IF boundary {decreases} the amplitude of $J_c$ by 1--2 orders of magnitude and, while the insulating barrier at the SF boundary shifts the $0$-$\pi$ transitions towards smaller values of $d_F$ ({solid black} lines), the additional N layer in the SINF part shifts it back to larger $d_F$.

This effect depends strongly on the value $\gamma_{NF}$, as can be seen from Figs.~\ref{fig:currents_plot_gNFchg}(f-h), where we show critical current densities $J_c (d_F)$ in the same FJJ configurations as in Figs.~\ref{fig:currents_plot_gNFchg}(b-d), but with fixed $d_N=0.4\xi_N$ and variable $\gamma_{NF}=1, \, 0.1, \, 0.01$. With decreasing $\gamma_{NF}$, the $0$-$\pi$ transitions get more shifted back to their positions without I layer. One may conclude that the thin N layer with small resistance ($\rho_N<\rho_F$) effectively ``smooths'' the order parameter in the SIF region.

For a physical explanation of this behavior one can imagine that a decrease of the amplitude of the superconducting pair wave-function in the F layer is connected to a decrease of the function $\theta_F$. In particular, the positions along the F layer where $\theta_F$ becomes zero correspond to sign reversals of the critical current and are therefore directly linked to the thicknesses $d_F$ where a $0$-$\pi$ transition occurs.

\begin{table}[t!]
\newcommand{\skiptab}{1.3cm}
\begin{tabular}{m{\skiptab}cccc}
Interface	& $\gamma_{BSF}$	& $\gamma_{BNF}$ & $\gamma_{BSN}\gamma_{NF}$ & Eq.~for $\chi^\pm$  \\ \hline
SF	& 0.001	& -- & -- & \eqref{eq:chi_r} \\
SIF	& 100	& -- & -- & \eqref{eq:chi_r} \\
SNF	& --	& 0.001 & 0.001 & \eqref{eq:chi_l} \\
SINF	& --	& 0.001 & 100 & \eqref{eq:chi_l}
\end{tabular}
\caption{{Parameters for the calculation of the critical current densities \eqref{eq:current:f:tot:res} shown in Fig.~\ref{fig:currents_plot_gNFchg}. The parameters $\gamma_B$ are responsible for the presence of an I layer, while the equation for the calculation of $\chi^\pm$ determines whether we consider an N layer or not. We keep the product $\gamma_{BSN}\gamma_{NF}$ constant because its outcome \eqref{eq:product:gamma} does not change during our analysis.}}
\label{tab:gamma_B}
\end{table}

This picture already helped to understand why an insulating layer at the SF interface shifts the $0$-$\pi$ transitions towards smaller values of $d_F$.\cite{Vasenko,Buzdin:PiJJ:JETPL2003} This is because the I layer induces a decreasing shift to $\theta_F$ at the SF interface, as can be seen from Eq.~\eqref{eq:boundary:fs} for $\gamma_{BSF}\gg1$. Since $\theta_F$ decreases monotonically from the interfaces into the F layer, this shift results in a shift of its zeros towards the interface. This in turn leads to a shift of the $0$-$\pi$ transitions to smaller $d_F$, as can be seen by comparing e.g. the black lines in Figs.~\ref{fig:currents_plot_gNFchg}(a) and \ref{fig:currents_plot_gNFchg}(b).

By inserting an N layer at the IF interface we can mitigate this effect. In fact, the function $\theta$ gets still decreased by the I layer, but the decrease of its derivative $\theta'$ may be less than compared to the case when the superconducting pair wave-function directly penetrates the F layer. This in turn leads to a shift of the $0$-$\pi$ transition back to larger $d_F$.

To explain this effect we replace the derivative $\theta_N'$ in Eq.~\eqref{eq:thetha_N^+:diff} with the help of Eq.~\eqref{eq:boundary:nf:diff} which leads us to the derivative
\begin{equation}
 \left[ \PartDeriv{\theta_F}{x} \right]_{x_{NF}}=\frac{\Omega d_N}{\xi_N \xi_F \gamma_{NF}}\sin{\theta_{N,{{S}}}}-\frac{\sin(\theta_S-\theta_{N,{{S}}})}{\gamma_{BSN}\gamma_{NF} \xi_F}
  \label{eq:thetha_F_NF:diff}
\end{equation}
at the F interface. For $d_N=0$, Eq.~\eqref{eq:thetha_F_NF:diff} resembles Eq.~\eqref{eq:boundary:fs}. Therefore we obtain, by using the values defined in Tab. \ref{tab:gamma_B}, the correct limiting results.

An increase of $d_N$ increases $\theta_F'$ and therefore shifts the $0$-$\pi$ transitions towards larger $d_F$, as shown by Figs.~\ref{fig:currents_plot_gNFchg}(b--d). Furthermore, from Eq.~\eqref{eq:thetha_F_NF:diff} can be understood why a smaller value of $\gamma_{NF}$ induces a larger increase of $\theta_F'$. This again shifts the $0$-$\pi$ transitions towards larger $d_F$, as shown by Figs.~\ref{fig:currents_plot_gNFchg}(f--h).

The same effect occurs in Fig.~\ref{fig:currents_plot_gNFchg}(e), but it has a different interpretation because the $0$-$\pi$ transitions are already shifted to large $d_F$ without an N layer, due to the absence of the I layer (black line). A small  value of $\gamma_{NF}$ does not change this situation significantly. However, if $\gamma_{NF}$ increases and therefore $\theta_F'$ decreases, the $0$-$\pi$ transitions get shifted to smaller $d_F$.

Note that these effects are related not only to the thickness of the N layer that may be small ($d_N \ll \xi_N$) but mainly to its conducting properties represented by $\gamma_{NF}$ \eqref{eq:gamma:bnf}.

\begin{figure}[t!]
\begin{center}
\includegraphics{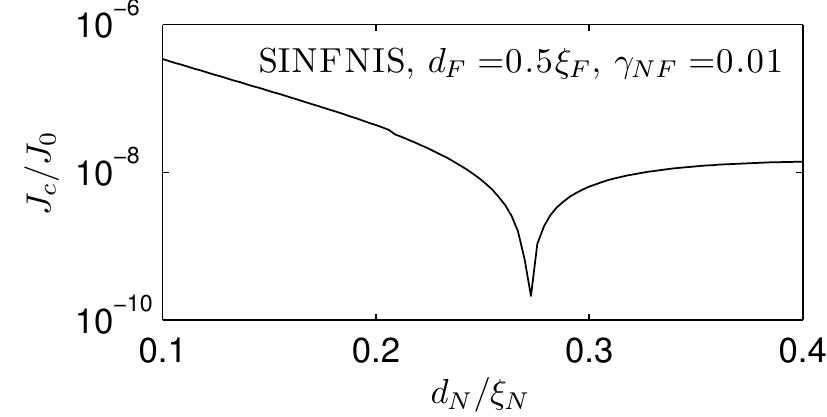}
  \caption{
  The critical current density $J_c(d_N)$~\eqref{eq:current:f:tot:res} in units of $J_0={\pi T_c}/{(\rho_F \xi_F e)}$ for FJJs in SINFNIS configuration. The F layer thickness $d_F=0.5\xi_F$ is constant, while the thickness $d_N$ of both N layers changes. In this way, we control the $0$-$\pi$-transition only by adjusting $d_N$. Analogous to Fig.~\ref{fig:currents_plot_gNFchg}(d), where the same FJJ configuration is analyzed for varying $d_F$, we use the suppression parameters of Tab. \ref{tab:gamma_B} and the parameters $h=30$, $T_c=9.2\units{K}$, $T=0. 5T_c$, $\eta=0$.
  }
  \label{fig:current_fct_of_d_N}
\end{center}
\end{figure}

{
The influence of N layers on FJJs can be seen most clearly when they are inserted at IF interfaces and $d_F$ is kept constant {not far from a 0-$\pi$ transition} while $d_N$ changes. In this way, \emph{the $0$-$\pi$ transition can be controlled by $d_N$}, as shown in Fig.~\ref{fig:current_fct_of_d_N}. Here we consider an SIFIS junction which is in the $0$ state for $d_F=0.5\xi_F$. By adding N layers at the IF interfaces and increasing their thicknesses simultaneously, we tune the FJJ into the $\pi$ regime. Figure \ref{fig:current_fct_of_d_N} considers the same FJJ configuration as Fig.~\ref{fig:currents_plot_gNFchg}(d), where $d_N$ is fixed and $d_F$ changes.
}

To understand the role of the boundary parameters in the $0$-$\pi$ transition patterns in more detail, it is useful to analyze it in a simple linear approximation. This approximation can be used if both S electrodes have nontransparent interfaces, or if $T \rightarrow T_c$. Then we may assume that $\theta\ll 1$, $G=\cos \theta \approx 1$, and $F \sim \sin \theta \approx \theta$. The general solution of the Usadel equations (\ref{eq:usadel:general}) in the non-superconducting layers has the form $\exp (\pm k_{N,F} x )$, where $k_N \equiv \sqrt{2\omega/D_N}$, $k_F \equiv \sqrt{2\tilde{\omega}/D_F} \equiv p+iq$, where $p$ and $q$ are real. The critical current density is given by the expression (\ref{eq:current:f:tot:def}). For FJJs without N layer, the critical current density was already calculated in Refs.~\onlinecite{Vasenko,BuzdinBaladie2003,faure:2006,vasenko:2011}.

\subsubsection{Transparent-interface structures: SFS, SNFS, SNFNS}

We start with the analysis of Figs. \ref{fig:currents_plot_gNFchg}(a) and \ref{fig:currents_plot_gNFchg}(e). For this purpose we assume that all interfaces are transparent, that is $\gamma_{BSF},\gamma_{BSN},{\gamma_{BNF}} \ll 1,$ and $T\rightarrow T_{c}.$ If $\gamma_{SF}\ll 1$, the critical current density of the SFS junction (cf. {solid black} lines) reads\cite{Buzdin:2005:Review}
\begin{equation}
J_{c} \sim \sum\limits_{\omega } \left[ \frac{\Delta ^{2}}{\omega ^{2}}\Real \frac{k_{F}}{\sinh (k_{F}d_{F})}\right]
\label{eq:Jc:SFS}
\end{equation}
and the positions of the $0$-$\pi $ transitions are defined by the solutions of the equation
\begin{equation}
\tan(qd_{F})=-\frac{p}{q}\tanh(pd_{F})
.
\end{equation}

This gives $qd_{F}\approx \pi -\arctan (p/q)$, and the first $0$-$\pi$ transition occurs at $\pi /2<$ $qd_{F}<\pi$. For a large exchange energy $ H\gg T_{c}$, we obtain $p\approx (1+\omega /2H)/\xi _{H}$ and $q\approx (1-\omega /2H)/\xi_{H}$. When we assume $p\approx q$, the first $0$-$\pi $ transition occurs at $d_{F}/\xi _{H}\approx 3\pi /4$, that is $d_F/\xi_F\approx 3\pi/\sqrt{8 h}\approx 0.6$, which is in good agreement with Figs. \ref{fig:currents_plot_gNFchg}(a) and \ref{fig:currents_plot_gNFchg}(e).

By adding normal layers in the case of $\gamma _{SN},\gamma _{NF}\ll 1$, we see that even for two extra layers in the SNFNS configuration, the critical current density
\begin{equation}
J_{c}\sim\sum\limits_{\omega } \left[ \frac{\Delta ^{2}}{\omega ^{2}}\frac{1}{\cosh ^{2}(k_{N}d_{N})}\Real  \frac{k_{F}}{\sinh (k_{F}d_{F})}\right]
\label{eq:Jc:SNFNS}
\end{equation}
differs not much from Eq.~(\ref{eq:Jc:SFS}). We only obtain an additional real factor $\cosh^{-2}(k_{N}d_{N})$, but the position of the $0$-$\pi $ transitions is still defined by the term marked as real part. Therefore, the positions of the $0$-$\pi $ transitions will be the same as in the SFS case, see Fig.~\ref{fig:currents_plot_gNFchg}(a) for one extra N layer. The small boundary parameter $\gamma_{SN}$ is needed in order to neglect the proximity effect in the S electrodes.

However, if $\gamma_{NF}=1$ in the SNFS junction ({dashed-}{dotted} line in Fig.~\ref{fig:currents_plot_gNFchg}(e)), the electrons may easily change between the N and F layers, since $\gamma_{NF}\sim \sqrt{D_F/D_N}$. Therefore, the Josephson phase drops partially along the N layer and the first $0$-$\pi $ transition shifts towards smaller values of $d_{F}$.

\subsubsection{Double-barrier structures SIFIS vs. SINFNIS}

In order to discuss the interplay of the N and I layers we jump to the description of the configurations shown by Figs. \ref{fig:currents_plot_gNFchg}(d) and \ref{fig:currents_plot_gNFchg}(h). Here the resistance of the insulating barriers is large $\gamma_{BSF},\gamma_{BSN} \gg 1,$ but the NF boundaries are still transparent ${\gamma_{BNF}}\ll 1,$ and we do not need any assumption about the temperature to use the linear approximation.

The critical current density of the SIFIS junction (cf. {solid black} lines) at $\gamma_{BSF}\gg 1$ is
\begin{equation}
J_{c}\sim\sum\limits_{\omega } \left[ \frac{\Delta ^{2}}{\gamma _{{BSF}}^{2}\xi _{F}^{2}\sqrt{\omega ^{2}+\Delta ^{2}}}\Real  \frac{1}{k_{F}\sinh (k_{F}d_{F})} \right] .
\label{eq:Jc:SIFIS}
\end{equation}

The points of the $0$-$\pi $ transitions are now defined by the
solutions of the equation
\begin{equation}
\tan(qd_{F})=\frac{p}{q}\tanh (pd_{F}).
\end{equation}
Here the assumption $p\approx q$ yields only $d_{F}=0$. At a large exchange energy $H\gg T_c$ the first $0$-$\pi $ transition occurs at $d_{F}/\xi _{H}<\pi /2$, that is $d_F/\xi_F< \pi/\sqrt{8 h}\approx 0.2$, which is in agreement with Figs. \ref{fig:currents_plot_gNFchg}(d) and \ref{fig:currents_plot_gNFchg}(h). Its exact position is defined by the factor $T/H$ as well as \cite{BuzdinBaladie2003} $\gamma_{BSF}$.

In the case of intermediate resistances $\gamma_{BSF} \sim 1$ of the SF interfaces of an SFS JJ\cite{Buzdin:2005:Review}, the critical current density reads
\begin{align}
&J_{c} \sim \sum\limits_{\omega } \left[ \frac{\Delta ^{2}}{\omega ^{2}} \times \right. \nonumber \\
& \left. \Real  \frac{k_{F}}{\sinh (k_{F}d_{F})(1+k_{F}^{2}\xi _{F}^{2}\Gamma
^{2})+2k_{F}\xi _{F}\Gamma \cosh (k_{F}d_{F})} \right] ,
\end{align}
which transforms into the two previous cases (\ref{eq:Jc:SFS}) and (\ref{eq:Jc:SIFIS}) for $\Gamma \equiv \gamma_{SF} \sqrt{\omega^{2}+\Delta ^{2}} / \left\vert \omega \right\vert \ll $ and $\gg 1$, respectively. The points of the $0$-$\pi $ transitions are defined by
\begin{equation}
\tan(qd_{F})=\frac{p(1+2\Gamma ^{2})\tanh (pd_{F})+4p\Gamma }{q(1-2\Gamma ^{2})}.
\end{equation}

If $2\Gamma >1$, that is $\gamma_{BSF}>\left\vert \pi T\right\vert /\sqrt{2(\pi ^{2}T^{2}+\Delta ^{2})}$, the first $0$-$\pi $ transition is located in the range $\pi /2<d_{F}/\xi _{H}<3\pi /4.$ If $\gamma_{BSF}<\left\vert \pi T\right\vert /\sqrt{2(\pi ^{2}T^{2}+\Delta ^{2})}$, it occurs at $0<d_{F}/\xi _{H}<\pi /2.$

In contrast, the critical current density of the SINFNIS junction at $\gamma_{BSN}\gg 1,$ at transparent NF interfaces ${\gamma_{BNF}}\ll 1$
and $\gamma_{NF}\ll 1$ , has the form
\begin{align}
J_{c}\ \sim \sum \limits_{\omega } \left[ \frac{\Delta ^{2}}{\sqrt{\omega^{2}+\Delta ^{2}}}\frac{1}{{\gamma_{BNF}^2}\xi_{N}^{2}k_{N}^{2}\sinh ^{2}(k_{N}d_{N})}\times \right. \nonumber \\
\Real \left. \frac{k_{F}}{\sinh(k_{F}d_{F})} \right] .
\end{align}
The $0$-$\pi $ transitions are defined by the zeros of the real part, which has the same form as in the case of SFS JJs with transparent interfaces (\ref{eq:Jc:SFS}). That is, the N layers have mitigated the effect of the I layers, which can be seen by comparing Fig.~\ref{fig:currents_plot_gNFchg}(d) with Fig.~\ref{fig:currents_plot_gNFchg}(a).

\subsubsection{SIFIS vs. SINFIS structures}

The effect of a single N layer on a double-barrier SIFIS junction, shown in Figs. \ref{fig:currents_plot_gNFchg}(c) and \ref{fig:currents_plot_gNFchg}(g), is discussed in the following. The critical current density of the SINFIS junction with the same boundary parameters as in the section before is given by
\begin{align}
J_{c} \sim \sum\limits_{\omega } \left[ \frac{\Delta ^{2}}{\sqrt{\omega
^{2}+\Delta ^{2}}}\frac{1}{{\gamma_{BNF}^2}\xi _{N}k_{N}\sinh
(k_{N}d_{N})} \times \right. \nonumber \\
\Real \left. \frac{k_{1}}{\cosh (k_{F}d_{F})} \right] .
\label{eq:Jc:SINFIS}
\end{align}

In this case the $0$-$\pi $ transitions are defined by the zeros of the function $\cos (qd_{F})$ and located at the positions where $d_{F}/\xi _{H}=\pi /2+\pi m,m=0,1,2...$ , that is they are also shifted towards larger $d_{F}$ in comparison with the ones of the SIFIS junction, see Figs.~\ref{fig:currents_plot_gNFchg}(c) and \ref{fig:currents_plot_gNFchg}(g).

In our previous article\cite{OurPRB2009} we obtained in fact the same expressions (\ref{eq:Jc:SIFIS}) and (\ref{eq:Jc:SINFIS}). There we assumed that the interface transparencies of both S electrodes are small, one of them due to the presence of an insulating barrier. In this way we analyzed SI$_{1}$FI$_{2}$S and SI$_{1}$NFI$_{2}$S structures with rather different transparencies of the I$_{1}$ and I$_{2}$ barriers. We found in the linear approximation that the critical current density for an SI$_{1}$NFI$_{2}$S FJJ is the same as the one for an SI$_{1}$FNI$_{2}$S structure.

\subsubsection{SIFS vs. SINFS structures}

If the structure contains only one insulating barrier, as in Figs. \ref{fig:currents_plot_gNFchg}(b) and \ref{fig:currents_plot_gNFchg}(f),  we may use the tunnel Hamiltonian method, which yields for the critical current density the expression
\begin{equation}
J_{c}\sim \sum_{\omega }\frac{\Delta ^{2}}{\sqrt{\omega^{2}+\Delta ^{2}}}\Real \left.\sin \theta _{N,{S}} \right.
.
\end{equation}

To use the linear approximation we shall assume that $T$ is close to $T_{c}$, and in order to neglect the proximity effect in the right S electrode we use the rigid boundary conditions $\gamma_{BSF},\gamma_{SF}\ll 1.$ We also
assume the N layer to be thin $d_N\ll \xi_N$. Then we obtain
\begin{equation}
\theta _{N,{S}}=\frac{1+\gamma_{NF}\frac{\xi _{F}k_{F}}{\sinh k_{F}d_{F}}
\left( \frac{d_{N}}{\xi _{N}}\cosh k_{F}d_{F}+\gamma_{BSN}\right) }{
1+\gamma_{NF}\frac{\xi _{F}k_{F}\cosh k_{F}d_{F}}{\sinh k_{F}d_{F}}\left(
\frac{d_{N}}{\xi _{N}}+\gamma_{BSN}\right) }
.
\end{equation}

To find the position of the first $0$-$\pi$ transition we assume $d_{F}\sim \xi _{H}$ and neglect $d_{N}/\xi _{N}\ll \gamma_{BSN}$,
because the last value is determined by the large resistance of the I barrier. The solution weakly depends on $d_{N}$ because the suppression of the superconducting correlation along the thin N layer is negligible in comparison with the one of the I barrier. However, the ratio of the N and F resistance, which defines via $\gamma_{NF}$ the derivative jump \eqref{eq:boundary:nf:diff} at the NF interface, still plays a role. Then the $0$-$\pi $ transition takes place at $d_F$, for which the equation
\begin{equation}
1+\gamma +2\gamma ^{2}\cos \frac{d_{F}}{\xi _{H}}+\gamma \left( \cos \frac{
d_{F}}{\xi _{H}}+\sin \frac{d_{F}}{\xi _{H}}\right) =0
\end{equation}
is satisfied.

If $\gamma \equiv \gamma_{BSN} \gamma_{NF} \xi _{F}/\xi _{H}\gg 1$, the main term gives $\cos (d_{F}/\xi _{H})=0$ and $d_{F}/\xi _{H}=\pi /2$, which corresponds to the solution for the SIFS FJJ \cite{Vasenko}. If $\gamma \gtrsim 1$ the position of the $0$-$\pi $ transition
shifts towards larger $d_{F}$ depending on $\gamma \sim \gamma_{NF}$, see Fig.~\ref{fig:currents_plot_gNFchg}(f). If $\gamma \ll 1$ we cannot use this approach assuming large $\gamma_{BSN}.$

\section{Comparison with experiment}\label{sec:comparison}

{To check our theory, we use data of SINFS experiments performed by M. Weides \emph{et al.}\cite{Weides:2006:SIFS-HiJcPiJJ} on Nb/Al$_2$O$_3$/Cu/Ni$_{0.6}$Cu$_{0.4}$/Nb FJJs.
}
{The samples used in these experiments include a 2 nm Cu interlayer between the I and F layer.}
{Using the same technology, new series of samples were produced, but the process was changed in order to delete the Cu layer. That is, we can compare SIFS and SINFS FJJs with the same layer properties including the concentration of the NiCu alloy. In Fig.~\ref{fig:experiment} we show {a fit of experimental data of critical current densities for different F layer thicknesses $d_F$ of both types of junctions. Dots correspond to  SIFS junctions} and triangles correspond to SINFS junctions.

{
We calculated the critical currents with the help of Eq.~\eqref{eq:current:f:tot:res}. In the case of the SIFS configuration we made use of Eq.~\eqref{eq:chi_r} to calculate the parameter $\chi^-$ and in the case of the SINFS configuration we used Eq.~\eqref{eq:chi_l}.
}

For our fit {we used the coherence lengths $\xi_N$=10 nm, $\xi_F$=7.60 nm and $\xi_H$=1.72 nm. Our exchange energy $H/k_B$=880 \units{K} is situated between the value 850\units{K} corresponding to the alloy Ni$_{0.53}$Cu$_{0.47}$\cite{Ryazanov:2001:SFS-PiJJ} and the value 930\units{K} of clean Ni.\cite{robinson:2006} The product $\tau_m H$=1/1.7 is similar to the one used by Weides \emph{et al.}\cite{Weides:2006:SIFS-HiJcPiJJ} Further values taken from this publication are the temperature T=4.2 K, junction area $A$=(100 \textmu m)$^2$ and resistivity $\rho_F$=54 \textmu$\Omega$cm. Additionally we used the damped critical temperature  T$_c$=7.2 K ͑of Nb and the resistivity $\rho_N$=0.66 \textmu$\Omega$cm.}

{
As we have shown in Fig.~\ref{fig:currents_plot_gNFchg}(f), a small suppression parameter $\gamma_{NF}<1$ results in a shift of the 0-$\pi$ transition to larger $d_F$ for the sample with N layer. This effect explains the shift of the $0$-$\pi$ transition} {observed in the experiments on SIFS and SINFS FJJs. The difference in the amplitude of the curves is attributed to the different thickness of the I barrier in these two sample series.}

This conclusion is supported by preliminary experimental observations on SIsFS junctions. These observations indicate that the introduction of a thin s interlayer, which should make a transition to the normal state if its thickness is of the order of the coherence length, shifts the $0$-$\pi$ transitions towards larger d$_F$.

\begin{figure}[t]
% \begin{center}
\includegraphics{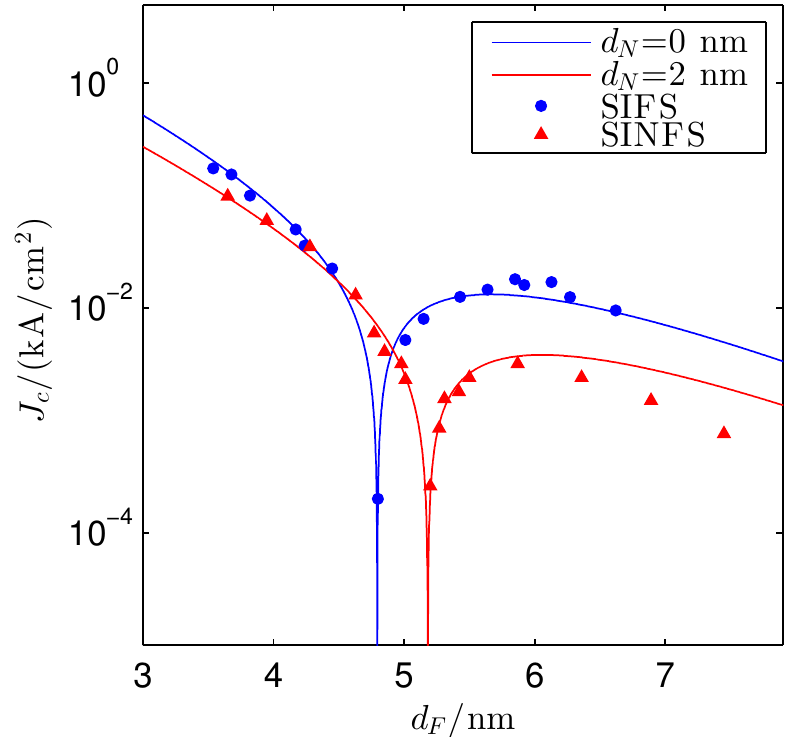}
  \caption{The critical current density \eqref{eq:current:f:tot:res} fitted to the experimental data of SIFS junctions { and SINFS junctions. For the calculation of $\chi^-$ in the SIFS case we used Eq.~\eqref{eq:chi_r} and in the SINFS case we used Eq.~\eqref{eq:chi_l}. Fitting parameters are $\gamma_{BSN}$=90000, $\gamma_{BNF}$=0.01, $\gamma_{BSF}$=0.1 and $\gamma_{NF}$=0.016.}}
  \label{fig:experiment}
% \end{center}
\end{figure}

\section{Conclusion}\label{sec:conclusion}

Using the Usadel equations we have calculated the critical current density of ferromagnetic Josephson junctions (FJJs) of different types, containing I and N layers at the SF interfaces and compared it to  critical current densities of structures without N layers. Such layers were technologically required in many FJJ experiments, but were not taken into account in previous models.

It was shown earlier \cite{Vasenko,BuzdinBaladie2003,faure:2006} that insulating barriers decrease the critical current density and shift the $0$-$\pi$ transitions to smaller values of the ferromagnet thickness $d_F$. A thin N layer inserted between S and I layers does not significantly influence the Josephson effect. However, if the N layer is inserted between I and F layers, it can have a large effect on the $J_c(d_F)$ curve. If additionally the transport properties of the F and N layers differ significantly ($\gamma_{NF} \ll 1$), the presence of the N layer shifts the first $0$-$\pi$ transition to larger $d_F$, see Figs.~\ref{fig:currents_plot_gNFchg}(b-d). {At certain values of $d_F$, the 0-$\pi$ transition can even be achieved by changing only $d_N$, see Fig.~\ref{fig:current_fct_of_d_N}.} {Finally, our theory allows the explanation of} { experimental data for SINFS and SIFS junctions, shown in Fig.~\ref{fig:experiment}.}

The oscillation period of $J_c(d_F)$ is still determined by the relation of the magnetic exchange energy $H$ and the diffusion coefficient $D_F$ in the dirty limit. At an average scattering strength this is in general not valid.\cite{OurPRB2011} If the transport
properties of the N layer between the I and F layer are the same as those of the ferromagnet, the $J_c(d_F)$  dependence does not change. This means in particular, that the dead layer\cite{RyazanovBuzdinJc(d)PRL2006,Blum:2002:IcOscillations,Sellier:2003:SFS,Weides:2006:SIFS-HiJcPiJJ,Weides:0-piLJJ,Pfeiffer:2008:SFS:semifluxon,Weides:2009} plays only a role if its properties differ from the ones of the ferromagnet, not only in terms of the absence of ferromagnetism, but also in terms of its resistance. The smaller the value of $\gamma_{NF}$, the larger is the change of the $J_c$ amplitude and the shift of the $0$-$\pi$ transitions, see Figs.~\ref{fig:currents_plot_gNFchg}(f-h).

The situation is completely different in the case of transparent SF interfaces, that is without an I layer in between. In this case the additional thin normal layer with conductivity much larger than the one of the ferromagnet ($\gamma_{NF}\ll1$) does not play any role. In the same setup, an N layer with transport properties similar to the ones of the ferromagnet ($\gamma_{NF}\approx1$) provides a shift of the $0$-$\pi$ transition to smaller $d_F$, see Fig.~\ref{fig:currents_plot_gNFchg}(e). {This process is explained in more detail after Eq.~\eqref{eq:thetha_F_NF:diff}.}

In summary, even a thin additional N layer may change the boundary conditions at the IF boundary depending on the
value of $\gamma_{NF}$. We conclude that it can effectively mitigate the effect of the insulating barrier on the decaying oscillations of the critical current density $J_c(d_F)$. Even technological thin N layers, which almost do not suppress the superconducting correlations, have to be taken into account for the explanation of experimental results concerning the Josephson effect in FJJs.

\acknowledgments
We thank Prof. V. V. Ryazanov for fruitful and stimulating discussions, N. G. P. thanks the CMPC RHUL for giving new ideas in stimulating discussions, and D. M. H. thanks Prof. W. P. Schleich and K. Vogel for giving him the possibility to work at the Lomonosov Moscow State University. Financial support by the DFG (Projects SFB/TRR-21 and KO 1953/11-1), the EPSRC (grant no. EP/J010618/1), the Russian Foundation for Basic Researches (RFBR grant no. 13-02-01452-a, 14-02-90018-Bel-a), and the Ministry of Education and Science of Russian Federation (grant no. 14Y26.31.0007) is gratefully acknowledged.

\appendix*

\section{N layer Green's function} \label{sec:deriv}

In this appendix we first show how to find the dependence of $\theta_{N,{F}}$ on ${\chi_{{N}}}\equiv \sin({\theta_{F,{N}}}/2)$ in order be able to solve Eq.~\eqref{eq:chi_l} numerically for $\chi_{{N}}$. Thereafter we reduce Eq.~\eqref{eq:chi_l} in the limiting case {$\eta,\gamma_{NF} \rightarrow 0$} to {an equation of fourth order in} $\chi_{{N}}$.

We start by solving the Usadel equation \eqref{eq:usadel:theta} in the case $j={N}$, that is
  \begin{equation}
    \xi_N^2\ \frac{\partial^2}{\partial x^2} \theta_N(x) = {\Omega} \sin \theta_N(x)
    , \label{eq:usadel:n}
  \end{equation}
where $\Omega \equiv \omega/(\pi T_c)$ because the exchange energy $h$ is zero in the N layer.

When we assume {$ \xi_N \gg d_N$}, the function $\theta_N(x)$ changes only slowly. Therefore, we make in the right-hand side of Eq.~\eqref{eq:usadel:n} the approximation
  \begin{equation}
    \sin \theta_N(x) \approx \sin \theta_{N,{{S}}} \equiv \mathrm{const}
    , \label{eq:approx:n}
  \end{equation}
where $\theta_{N,{{S}}}\equiv \theta_N(x_{SN})$. Note that we cannot neglect this term because $\theta_N(x)$ may be of the order of $\theta_S$, depending on the boundary parameters. The solution of Eq.~\eqref{eq:usadel:n} using the approximation \eqref{eq:approx:n} reads
  \begin{equation}
   \theta_N(x)=\frac{\Omega}{2\xi_N^2}\sin\theta_{N,{{S}}}\ (x-x_{SN})^2+a(x-x_{SN})+\theta_{N,{{S}}}
   . \label{eq:sol:n}
  \end{equation}

Inserting the constant
  \begin{equation}
   a\ = \frac{1}{\gamma_{BSN}\ \xi_N} \sin(\theta_{N,{{S}}}-\theta_S)
   , \label{eq:a:def}
  \end{equation}
determined from the the boundary condition \eqref{eq:boundary:sn} at the SN interface, into the Green's function \eqref{eq:sol:n} at the position $x_{NF}$ connects the {NF} boundary value
\begin{equation}
 {\theta_{N,{{F}}}}=\frac{\Omega d_N^2}{2\xi_N^2}\sin{\theta_{N,{{S}}}}+\frac{d_N}{\gamma_{BSN}\ \xi_N} \sin(\theta_{N,{{S}}}-\theta_S)+{\theta_{N,{{S}}}}
 \label{eq:thetha_N^+}
\end{equation}
to the {SN} boundary value $\theta_{N,{{S}}}$, which we determine in the next step.

For this purpose we use {the integrated sine-Gordon equation \eqref{eq:usadel:theta:integrated} at the position $x_{NF}$ and insert it into the differentiability condition \eqref{eq:boundary:nf:diff} to obtain}
\begin{equation}
 {
 -2\gamma_{NF} \sqrt{\widetilde \Omega +\eta \cos^2\frac{\theta_{F,{N}}}{2}} \sin\frac{{\theta_{F,N}}}{2} = \xi_N \left[ {\PartDeriv{}{x} \theta_N} \right]_{x_{NF}}
 }
 . \label{eq:boundary:nf:deri}
\end{equation}
{Here we replace the right-hand side by} the derivative
\begin{equation}
 \left[ \PartDeriv{\theta_N}{x} \right]_{x_{NF}}=\frac{\Omega d_N}{\xi_N^2}\sin{\theta_{N,{{S}}}}+\frac{\sin(\theta_{N,{{S}}}-\theta_S)}{\gamma_{BSN}\ \xi_N}
 \label{eq:thetha_N^+:diff}
\end{equation}
of the function $\theta_N(x)$ from Eq.~\eqref{eq:sol:n}. 

{These steps lead} us with the definition ${\chi_{{N}}} \equiv \sin({\theta_{F,{{N}}}}/2)$ to
\begin{align}
 {
 - 2{\gamma_{NF}}{\gamma_{BSN}} \sqrt{\widetilde \Omega + \eta(1-\chi_N^2)}\ \chi_N
  }
  \nonumber \\
  {
  =  \Omega \frac{d_N}{\xi_N} \gamma_{BSN} \sin{\theta_{N,S}}
  + \sin(\theta_{N,S}-\theta_S)
  }
 . \label{eq:sin:theta_N^-}
\end{align}

This equation can be written as an equation of second order in $\mu \equiv \sin{\theta_{N,{{S}}}}$ and can therefore be solved exactly for $\theta_{N,{{S}}}$. Inserting the result into Eq.~\eqref{eq:thetha_N^+} gives us $\theta_{N,{{F}}}$ as a function of $\chi_{{N}}$ which itself, when inserted into Eq.~\eqref{eq:chi_l}, allows us to determine finally $\chi_{{N}}$ by solving the transcendental equation \eqref{eq:chi_l} numerically.

In the following we consider the {limit $\eta,\gamma_{NF} \rightarrow 0$} to reduce Eq.~\eqref{eq:chi_l} to an equation of fourth order in $\chi_{{N}}$. This {limit} allows us to neglect the term containing $\chi_{{N}}$ in Eq.~\eqref{eq:sin:theta_N^-}. Together with the definition \eqref{eq:a:def} we obtain the equation
\begin{equation}
 \sin{\theta_{N,{{S}}}}=-\frac{\xi_N^2}{\Omega d_N}{a}
 , \label{eq:sin:theta_N^-:shorter}
\end{equation}
which we use to replace ${\theta_{N,{{S}}}}$ in Eq.~\eqref{eq:a:def}.

Solving the resulting equation for ${a}$ and re-inserting it into Eq.~\eqref{eq:sin:theta_N^-:shorter} leads us to the expression
\begin{align}
 \sin{\theta_{N,{{S}}}}=\lambda \sin \theta_S,
 \label{eq:theta_Nl}
\end{align}
where we used the definition
\begin{align}
 \lambda \equiv \left(1+2\cos \theta_S\gamma_{BSN}\frac{\Omega d_N}{\xi_N}+\gamma_{BSN}^2\frac{\Omega^2 d_N^2}{\xi_N^2}\right)^{-1/2}
 . \label{eq:sin:theta_N^-:finally}
\end{align}
With the help of Eq.~\eqref{eq:theta_Nl} we replace $\theta_{N,{{S}}}$ in Eq.~\eqref{eq:thetha_N^+}, which in turn is used in Eq.~\eqref{eq:chi_l} to reduce it finally {together with $\eta \rightarrow 0$} to an equation of fourth order in ${\chi_{{N}}}$.

% \bibliography{snfs}

%

\end{document}